\documentclass[twoside]{article}
\usepackage{qic}

\usepackage{amsmath,amssymb,latexsym}
\usepackage{graphicx}
\usepackage[usenames]{color}
\usepackage[frame,line,arrow,matrix,tips,color]{xy}
\usepackage{subfigure}
\usepackage{epsfig}

\usepackage{hyperref}

\newcommand \SWAP {{\tt SWAP}}
\newcommand \CNOT {{\tt CNOT}}

\newcommand \CZ {{\tt CZ}}

\newcommand\beq{\begin{equation}}
\newcommand\eeq{\end{equation}}
\newcommand\bea{\begin{eqnarray}}
\newcommand\eea{\end{eqnarray}}

\newcommand{\ba}{\begin{array}}
\newcommand{\ea}{\end{array}}

\include{Qcircuit}

\begin{document}
\setlength{\textheight}{8.0truein}    

\runninghead{Noise Threshold for a Fault-Tolerant Two-Dimensional Lattice Architecture}
            {K.~M.~Svore, D.~P.~DiVincenzo, B.~M.~Terhal}

\normalsize\textlineskip
\thispagestyle{empty}
\setcounter{page}{1}

\copyrightheading{0}{0}{2003}{000--000}

\vspace*{0.88truein}

\alphfootnote

\fpage{1}

\centerline{\bf Noise Threshold for a Fault-Tolerant
Two-Dimensional Lattice Architecture} \vspace*{0.035truein}
\vspace*{0.37truein} \centerline{\footnotesize Krysta
M.~Svore\footnote{kmsvore@cs.columbia.edu}} \vspace*{0.015truein}
\centerline{\footnotesize\it Dept. of Computer Science, Columbia
University, 1214 Amsterdam Ave., MC:0401} \baselineskip=10pt
\centerline{\footnotesize\it New York, NY 10027,USA}
\vspace*{10pt} \centerline{\footnotesize David
P.~DiVincenzo\footnote{divince@watson.ibm.com}}
\vspace*{0.015truein} \centerline{\footnotesize\it IBM Research,
PO Box 218,} \baselineskip=10pt \centerline{\footnotesize\it
Yorktown Heights, NY 10570 USA} \vspace*{10pt}
\centerline{\footnotesize Barbara
M.~Terhal\footnote{bterhal@gmail.com}} \vspace*{0.015truein}
\centerline{\footnotesize\it IBM Research, PO Box 218,}
\baselineskip=10pt \centerline{\footnotesize\it Yorktown Heights,
NY 10570 USA} \vspace*{0.225truein} \publisher{(received
date)}{(revised date)}

\vspace*{0.21truein}

\abstracts{
We consider a model of quantum computation in which the set of
operations is limited to nearest-neighbor interactions on a 2D
lattice. We model movement of qubits with noisy \SWAP\ operations.
For this architecture we design a fault-tolerant coding scheme using
the concatenated $[[7,1,3]]$ Steane code. Our scheme is potentially
applicable to ion-trap and solid-state quantum technologies. We
calculate a lower bound on the noise threshold for our local model
using a detailed failure probability analysis. We obtain a threshold
of $1.85 \times 10^{-5}$ for the local setting, where memory error rates
are one-tenth of the failure rates of gates, measurement, and
preparation steps. For the analogous nonlocal setting, we obtain a
noise threshold of $3.61 \times 10^{-5}$. Our results thus show that the
additional \SWAP\ operations required to move qubits in the local
model affect the noise threshold only moderately.
}{}{}

\vspace*{10pt}

\vspace*{3pt}
\communicate{to be filled by the Editorial}

\vspace*{1pt}\textlineskip    

\section{Introduction}
In order to perform large-scale quantum computation reliably, any
quantum computer technology will require error correction.
Error-correcting quantum data is useful if the logical noise level
is lowered by coding and error-correcting. Since error correction is
noisy, a decrease in logical noise level only occurs for
sufficiently low initial noise rates, below a certain {\em noise
threshold} \cite{AB:faulttol,KLZ:faulttol,preskill:faulttol,TB:ft} .
It is clear that it is of great importance for the future of quantum
computation to determine the values of such noise thresholds since
they set the level of accuracy to be achieved by experimental
programs.

Over the past 10 years estimates for noise thresholds for various
{\em code architectures} \cite{steane:overhead, reichardt:ft,
knill:nature, STD:localft} have been determined. These estimates,
depicted in the overview in Fig.~1 of \cite{SCCA:flowmap}, have
varying degrees of accuracy and rigor. In Ref.~\cite{AGP:ft},
entirely rigorous criteria for fault-tolerant design and threshold
analysis were formulated. These rigorous criteria allow, at least
for small error-correcting codes, an accurate determination of the
noise threshold for simple stochastic noise models.


In many threshold analyses it has been assumed that interactions
between qubits can be performed in a manner independent of
their spatial separation. In reality, interactions between
stationary qubits are typically restricted to occur between nearest
neighbors in some low-dimensional spatial layout, the {\em spatial
architecture}. A transport mechanism will thus be required to bring
qubits together. An example is the scalable ion-trap architecture
where ions are shuttled between interconnected ion traps
\cite{KMW:iontrap}. Transport mechanisms will also be required in
solid-state implementations, such as the method involving electron
spins in quantum dots \cite{petta+:science}, phosphorus spin qubits
in silicon \cite{SDK:silicon}, Josephson-junction qubits
\cite{koch+:jj} or neutral atoms in a 2D optical lattice \cite{SAJP:coldatoms}.

Previous work \cite{STD:localft, AB:ftsiam, gottesman:localft} has
established the existence of a noise threshold in a local model. In
Ref.~\cite{STD:localft} we made a first attempt to also estimate the
effect of locality on the noise threshold, in particular the
dependence on a code-dependent scale parameter and the failure rates
of qubit transportation. In Ref.~\cite{SBF+:localft}, the authors
derive a noise threshold of $O(10^{-7})$ for the [[7,1,3]] code for a
quasi-one-dimensional architecture. Some thresholds related to
ion-trap systems are also analyzed in Refs.~\cite{MTC+:ionarch,metodi+:ionarch2}.
For spin-qubits, Refs.~\cite{hollen+:ft2d} and \cite{taylor+:ftnature} estimate noise
thresholds taking into account transport mechanisms in conjunction
with explicit two-dimensional layouts. None of these papers employ
the rigorous methods of analysis outlined in Ref.~\cite{AGP:ft}. We
believe the use of rigorous methods can considerably impact the threshold values that
are found. We briefly argue in Section \ref{sec:why10-5} why
the methodology laid out in Ref. \cite{AGP:ft} gives fairly tight bounds
for the fault-tolerance threshold. This implies that some of the
previous threshold estimates for the [[7,1,3]] code have been too
optimistic.

In this paper we consider a two-dimensional nearest-neighbor lattice
architecture using logical noisy \SWAP\ gates as the basic qubit
transport mechanism. It is clear that any viable layout for
stationary qubits has to be in a two-dimensional plane so that
classical control fields can access the qubits from the third
dimension. In the case of a quasi-1D or self-similar architecture
(such as the H-tree), the classical controls could also lie in the
two-dimensional plane itself\footnote{A purely one-dimensional
fault-tolerant architecture is impossible if one uses a distance-3
code. The reason is that such an architecture necessitates swapping
data qubits inside a block which may generate a two-qubit error due
to one failed {\tt SWAP}. For a distance-3 code, such errors cannot
be corrected. This implies that the level-1 error rate can never be
smaller than the base error rate. A purely one-dimensional
fault-tolerant architecture is not excluded for higher distance
codes, but the threshold noise value may be very poor.}. In our
analysis, we consider the simplest possible noise model, namely
independent adversarial stochastic noise. In this noise model, each
location $\ell$ independently undergoes an error with some
probability $\gamma_\ell$. The error that occurs can be chosen
adversarially so that in our threshold analysis we assume that for
every location the worst among the three Pauli errors {\tt X}, {\tt
Y}, {\tt Z} occurs.

We find, somewhat surprisingly, that the threshold noise value for
our 2D lattice architecture is close to the threshold noise value
for a nonlocal architecture --- both are $O(10^{-5})$.  We
understand this small dependence on locality to arise from (1) the
dominance of the noisy two-qubit gates and (2) the fact that we have
optimized the circuits to be space-efficient in a dense, 2D square
lattice, with a surprisingly small cell size of $6\times 8$ for the
[[7,1,3]] code, resulting in a small amount of moving.

Given the good performance of our {\tt SWAP}-based architecture, we
believe that it is unnecessary to seek alternative means of
transportation (teleportation and entanglement purification for
example), at least for the [[7,1,3]] code in this 2D layout. As the
reader will observe, the noisy {\tt SWAP} gates simply
model any physical scheme in which qubits are transported through
some quantum channel. In our model, this channel will have a fixed
time-delay and a noise level that depends on the choice of noise
level for the {\tt SWAP}s. It will also be clear from our analysis
that no really long-distance means of transportation is needed for
fault-tolerant quantum computation.


\subsection{A Local 2D Lattice Architecture}\label{sec:lattice}
\noindent
Let us assume that we have modified an unencoded circuit $M$ so that
interactions among the qubits are nearest-neighbor on a 2D lattice.
This requires adding \SWAP\ gates in the unencoded circuit, which
can be implemented using three controlled-NOT ({\tt CNOT}) gates. We
will also be using \SWAP\ gates in our encoded circuit, but we will
treat these gates as elementary (and not composite) gates.

After one level of encoding using Steane's [[7,1,3]] code, every
qubit in $M$ is replaced by a 6$\times$8 cell of qubits, 48 qubits
in total, as depicted in Fig.~\ref{fig:lattice}. In a concatenated
architecture, the next level of encoding is obtained by replacing
each qubit in the cell again by a 6$\times$8 cell, and so on. In
this way, two-qubit {\tt CNOT} gates between nearest-neighbor
qubits in the cell become (transversal) \CNOT\ gates between two
neighboring cells. The cell functions as (1) the spatial unit in
which error correction takes place, (2) the `resting place' for a
block of data qubits, and (3) the space to create a special type
of ancillary encoded state $\ket{\overline{a}_{\theta}}$ (see
Sections \ref{sec:local1prep} and \ref{sec:inj}) that is used in
the computation.

For error correction, the 48 qubits of the cell are divided into
three sets, see Fig.~\ref{fig:lattice}.  There are seven physical
data qubits labeled by $d$ in the figure. In the snapshot, the
data qubits sit at a special location in the cell that we refer to
as {\em home base}. When the data qubits interact with ancilla
qubits for syndrome extraction, the data qubits will always be
sitting at home base. The cell also contains at most ten ancilla
qubits at a time. Three of these ancilla qubits are verification
qubits, labeled by $v$ in the figure. The other seven are the
proper ancilla qubits (labeled by $a$) that will receive the error
syndrome. The rest of the cell is filled by dummy qubits (labeled
by $O$). The state of the dummy qubits is irrelevant for the
computation; they are only used for swapping with data or ancilla
qubits. One can view the dummy qubits as forming channels through
which the other qubits can pass. Since ancilla qubits remain local
to the cell and only data qubits interact with data qubits in
neighboring cells, we place data qubits on the exterior of the
cell and devote the interior region of the lattice cell to ancilla
preparation and verification for error correction.

Interactions between data qubits in neighboring cells can occur
between two lattice cells situated on a horizontal axis, i.e., data
qubits must move horizontally from home base to interact, or on a
vertical axis, i.e., data qubits must move vertically from home base
to interact. We include both a vertical and a horizontal
transportation channel, consisting of dummy qubits, to be used for
horizontal and vertical transportation of data qubits. The channels
occupy the first row and first column of the lattice cell and they can be used by
qubits in neighboring cells as well (see the discussion in Section \ref{sec:spacetime}).

We assume \SWAP\ to be our mechanism for qubit transport, where a
qubit moves by swapping with an adjacent qubit.  To guarantee the
fault tolerance of each \SWAP\ operation we only {\em swap a data
or ancilla qubit with a dummy qubit}. Thus, in the layout shown in
Fig.~\ref{fig:lattice}, we alternate dummy qubits with data or
ancilla qubits to allow for proper fault-tolerant movement.

\begin{figure}[htpb]
\centerline{ \mbox{ {\scalebox{1.0}{
\xymatrix@=30pt@!0{\textcolor{Orange}{O}  & \textcolor{Orange}{O}
& \textcolor{Orange}{O}  & \textcolor{Orange}{O}  &
\textcolor{Orange}{O}  &
\textcolor{Orange}{O}  & \textcolor{Orange}{O}  & \textcolor{Orange}{O} \\
\textcolor{Orange}{O}  & \textcolor{Red}{d6}  &
\textcolor{Black}{O} & \textcolor{Red}{d5}  & \textcolor{Black}{O}
& \textcolor{Red}{d3}  & \textcolor{Black}{O}  &
\textcolor{Black}{O}
\\
\textcolor{Orange}{O}  & \textcolor{Black}{O}  &
\textcolor{Blue}{v3}  & \textcolor{Green}{a3}  &
\textcolor{Green}{a6} \ar@{->}[d] & \textcolor{Black}{O}  &
\textcolor{Black}{O}  & \textcolor{Black}{O}  \\
\textcolor{Orange}{O}  & \textcolor{Black}{O}  &
\textcolor{Blue}{v2} \ar@{->}[u] & \textcolor{Green}{a5}
\ar@{->}[u] & \textcolor{Green}{a4}  & \textcolor{Green}{a1}
\ar@2{<=>}[u] & \textcolor{Black}{O}  & \textcolor{Black}{O} \\
\textcolor{Orange}{O}  & \textcolor{Black}{O}  &
\textcolor{Blue}{P_z(v1)}  & \textcolor{Green}{a2}  &
\textcolor{Black}{O}  & \textcolor{Green}{a7} \ar@2{<=>}[l] &
\textcolor{Black}{O}  & \textcolor{Black}{O} \\
\textcolor{Orange}{O}  & \textcolor{Red}{d4}  &
\textcolor{Black}{O} & \textcolor{Red}{d2}  & \textcolor{Black}{O}
& \textcolor{Red}{d1}  & \textcolor{Black}{O}  &
\textcolor{Red}{d7} \POS"3,2"."5,2"."3,8"."5,8"!C*+<1.2em>\frm{--}
\POS"1,1"."6,1"."1,8"."6,8"!C*+<2.0em>\frm{-}}}}}} \vspace*{13pt}
\fcaption{\label{fig:lattice} Snapshot of the operations and state
of a 6$\times$8 lattice cell during the ancilla preparation part
of error correction. Ancilla qubits are labeled with
$\textcolor{Green}{a}$ and data qubits are labeled with
$\textcolor{Red}{d}$. Ancilla qubits labeled with
$\textcolor{Blue}{v}$ are used for verification. $O$ locations
represent dummy qubits that are used for swapping operations. The
inner dashed region is the area used for ancilla preparation and
verification. The vertical and horizontal dummy qubit border
regions represent channels (labeled as $\textcolor{Orange}{O}$)
used for qubit transportation. Single arrows represent \CNOT\
gates, double arrows represent \SWAP\ operations, and $P_z$
represents preparation of a $|0\rangle$ state.}
\end{figure}


\section{Review of Definitions and Analysis Method for the [[7,1,3]] Code}
\label{sec:review} \noindent We now review some definitions and
properties of the standard fault-tolerant analysis for the
[[7,1,3]] code; this has been described in detail in
Ref.~\cite{AGP:ft}. For the [[7,1,3]] code, the Clifford-group
gates (two-qubit {\tt CNOT}, single-qubit Hadamard ({\tt H}), and
single-qubit phase gate ({\tt S}), i.e., $\ket{0}\bra{0}+i
\ket{1}\bra{1}$) are transversal. In order to make the gate set
universal, one needs a non-Clifford-group gate; we choose the
single-qubit gate ${\tt T}=e^{-i \pi {\tt Z}/8}$.  We believe that
this choice for a non-Clifford-group gate is better than trying to
implement, by local means, a three-qubit Toffoli gate.

\begin{figure}[htb]
\centerline{
\includegraphics{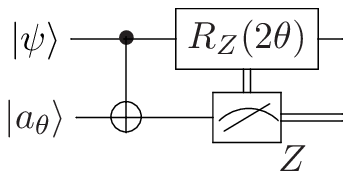}}
\vspace*{13pt}
\fcaption{\label{fig:genST}Circuit for implementing the single-qubit gate $R_{\tt Z}(\theta)$. If the measurement result is $\ket{1}$, then we correct the state by applying $R_{\tt Z}(2 \theta)$.  The output state is $R_{\tt Z}(\theta)\ket{\psi}$.}
\end{figure}

For reasons discussed later, we choose to implement the gate {\tt
T}, as well as that gate {\tt S}, as in Fig.~\ref{fig:genST}. Note
that ${\tt T}=R_{\tt Z}(\pi/4)$ and ${\tt S}=R_{\tt Z}(\pi/2)$,
where $R_Z(\theta) = e^{-i \frac{\theta}{2}{\tt Z}}$. In these
circuits, the state $\ket{a_{\theta}} = R_{\tt Z}(\theta)\ket{+}$.
The corrective operation $R_{\tt Z}(2\theta)$ for the {\tt T} gate
is the {\tt S} gate and the corrective operation for the {\tt S}
gate is {\tt Z}. We emphasize that with this construction, the
only gates in the {\em unencoded} circuit are {\tt X}, {\tt Y},
{\tt Z}, {\tt H}, and \CNOT\ (in addition to state preparations
and measurements).

{\em Locations} in a quantum circuit are defined to be gates,
single-qubit state preparations, measurement steps, or memory (wait)
locations; they can all be executed in a single timestep. After one
level of encoding, every location (denoted as 0-Ga) is mapped onto a
1-rectangle (1-Rec), a space-time region in the encoded circuit,
which consists of the encoded gate (1-Ga) followed by
error correction (1-EC), as shown in Fig.~\ref{fig:1Rec}. For transversal
gates, the 1-Ga consists of performing the 0-Gas on each qubit in
the block(s). The encoded version of a preparation location is
called a 1-Prep. For the [[7,1,3]] code, it consists of at most two
attempts at preparing the desired encoded state; each attempt is
followed by a verification step. If the verification of the first
preparation fails, we repeat the preparation and verification. If it
fails again, we continue with the failed encoded state which may
contain two errors. To make a 1-Rec, we follow the 1-Prep by a 1-EC.
The encoded version of a measurement location is called a 1-Meas. In
the $[[7,1,3]]$ code, it is implemented as seven parallel
measurements. In our model, we consider both measurement in the {\tt X} basis
and measurement in the {\tt Z} basis to be single locations (rather than
converting {\tt X} measurement into {\tt Z} measurement using a {\tt
H} gate).  For the measurement 1-Rec, the 1-EC is simply the
classical processing of the measurement outcomes.

\begin{figure}[htbp]
\centerline{
\mbox{ \Qcircuit @C=1em @R=.5em {
      & \multigate{6}{\mbox{1-Ga}} &
\multigate{6}{\mbox{1-EC}} &
\qw \\
    & \ghost{1-Ga} &
\ghost{1-EC} &
\qw \\
    & \ghost{1-Ga} &
\ghost{1-EC} & \qw
\\ & \ghost{1-Ga} &
\ghost{1-EC} & \qw \\
    & \ghost{1-Ga} &
\ghost{1-EC}&
\qw \\
    & \ghost{1-Ga} &
\ghost{1-EC} &
\qw \\
    & \ghost{1-Ga} &
\ghost{1-EC}& \qw \gategroup{1}{2}{7}{3}{.6em}{--} }} }
\vspace*{13pt}
\fcaption{\label{fig:1Rec} A 1-rectangle (1-Rec), indicated by a dashed box, that replaces a single-qubit 0-Ga location. The 1-Rec consists of the encoded fault-tolerant implementation of the 0-Ga (1-Ga) followed by an error-correction procedure (1-EC).}
\end{figure}

For the fault-tolerance analysis, one also defines an extended 1-Rec
(1-exRec) which consists of a 1-Rec with its preceding 1-EC(s) on the
input block(s). The 1-exRec for a preparation or measurement
location coincides with the 1-Rec for preparation or measurement.

In order to be fault-tolerant, the 1-EC, 1-Ga, and 1-Prep must satisfy
the fault-tolerance requirements listed in Ref.~\cite{AGP:ft} and
reproduced in Appendix \ref{app:ftprop}. Our constructions obey
these requirements with the caveat that the 1-EC does not satisfy
the second property. As we will show, it can leave an {\tt X} error
and a {\tt Z} error on the data after its application. Thus for
fault tolerance, it is necessary that the 1-Gas never transform a
{\tt X} or {\tt Z} error into a {\tt Y} error since this would
correspond to having two {\tt X} (or {\tt Z}) errors in a block.
This is achieved by only having 0-Gas which are {\tt X}, {\tt Y},
{\tt Z}, {\tt H}, and {\tt CNOT} (and not {\tt S}).

If the fault-tolerance requirements are fulfilled, it can be shown
that if a 1-exRec is good --- for a distance-3 code, such as the
[[7,1,3]] code, this means that the 1-exRec {\em contains at most
one fault}--- then the dynamics induced by the 1-Rec contained in
the 1-exRec is correct \cite{AGP:ft}, as defined in this picture:
\begin{center}
\begin{picture}(292,24)
\put(0,12){\line(1,0){10}}
\put(10,0){\framebox(48,24){\shortstack{correct\\$1$-Rec}}}
\put(58,12){\line(1,0){10}}
\put(68,0){\framebox(48,24){\shortstack{perfect\\$1$-decoder}}}
\put(116,12){\line(1,0){10}}
\put(126,6){\makebox(20,12){=}}
\put(146,12){\line(1,0){10}}
\put(156,0){\framebox(48,24){\shortstack{perfect\\$1$-decoder}}}
\put(204,12){\line(1,0){10}}
\put(214,0){\framebox(48,24){\shortstack{ideal\\$0$-Ga}}}
\put(262,12){\line(1,0){10}}
\put(272,6){\makebox(20,12){.}}
\end{picture}
\end{center}

A 1-exRec containing two faults can be bad, but not all higher-order
faults lead to two errors on the data. In order to give an accurate
estimate of the threshold, Ref.~\cite{AGP:ft} introduced the notion
of benign and malignant fault patterns. A set of locations is {\em
benign} in a 1-exRec if the 1-Rec contained in the 1-exRec is
correct. Otherwise, the set of locations is called malignant. With
this definition one can redefine a bad or {\em failed} 1-exRec,
namely, as a 1-exRec that contains faults at malignant locations. In
our analysis, we assume any pattern of three faults or more
is always malignant. We discuss this assumption in Section
\ref{sec:calcfail} since it has a small but non-zero effect on the
threshold estimate.

We denote the failure probability of a location of type $\ell$ as
$\gamma_\ell \equiv \gamma_\ell^0$. The failure probability at
encoding level $r$ for location $\ell$ can be computed with a
multi-dimensional map ${\bf F}$, with ${\bf F}_\ell({\bf
\gamma}^{r-1}) = \gamma_\ell^r$. The map ${\bf F}$ is independent of
concatenation level and is determined by the number of malignant and
higher-order errors in the 1-exRec for the different locations. An
initial vector of probabilities $\gamma^0$ is said to be below the
threshold
if there is a concatenation level
$r_{\epsilon}$ such that \beq \forall \epsilon, \exists
r_{\epsilon}, \forall
\ell,\;\;\gamma_{\ell}^{r_{\epsilon}}=\left({\bf
F}^{r_{\epsilon}}(\gamma^0)\right)_{\ell} \leq \epsilon. \eeq


In Sections \ref{sec:method} and \ref{sec:results}, we describe how
we calculate this map and the threshold results that we obtain for
the 2D local and the nonlocal model. We now describe the
local fault-tolerant circuits that we use in our analysis.

\section{Local Fault-tolerant Circuitry}
\label{sec:localft}
\noindent
Before we describe our error-correction circuitry, we first
describe local 1-exRecs.
In a local architecture, a 1-Ga which replaces a two-qubit 0-Ga
includes the necessary \SWAP\ operations to move the two encoded
qubits together. At the next level of concatenation, each new
location, including a {\tt SWAP}, is replaced by its 1-Rec. Our
\SWAP\ gates involve one dummy qubit and one data qubit; it is
unnecessary to do error correction on the dummy qubit. Therefore,
the \SWAP\ 1-Rec effectively behaves as a 1-Rec for a single-qubit
gate since it requires a 1-EC on only one block of qubits.

\subsection{Local Steane Error Correction (1-EC)}\label{sec:localec}
\noindent
For $[[7,1,3]]$ error correction, we have a choice between (1)
Shor-type error correction, where the syndrome for each stabilizer
operator is measured separately, (2) Steane-type error correction,
where the entire {\tt X}- or {\tt Z}-error syndrome is copied onto a
verified ancilla block, or (3) Knill-type error correction, where the
data qubits are teleported into a new block and {\tt X}- and {\tt
Z}-error-syndrome information is obtained via a Bell measurement.
For our local layout, it is important to minimize the use of space
and thus movement, potentially at the cost of increased waiting and
memory errors. The reason is that we assume memory errors will
typically occur at lower rates than \SWAP\ errors. This optimization
would seem to point to a sequential execution of the most
space-efficient error correction, which is Shor's. However, in
Shor-type error correction, the number of gates between data and
ancilla qubits is more than in Steane-type error correction, which can
negatively affect the threshold.

Therefore, we opt for a variant of Steane's error correction,
schematically drawn in Fig.~\ref{fig:Steane_QEC}. We use a {\em
deterministic} error-correction network, in which we sequentially
prepare three ancilla blocks in the encoded 0-state $\ket{\bar{0}}$.
If a single failure causes one ancilla block to fail verification,
then two ancilla blocks remain, one each for {\tt X}- and {\tt
Z}-error correction. If two ancilla blocks fail verification, which
by construction can only happen when two faults occur, we cannot
perform full error correction and we count that as a failure of the
data in this rectangle.
If the first two ancilla blocks pass verification, then there is no
need to prepare a third ancilla block.
Instead, the data block waits for the same number of timesteps that would have been required for the third
ancilla block preparation.

\begin{figure}[hb]
\centerline{ \mbox{\Qcircuit @C=1em @R=.7em {
&&&&&&&&&&\\
&\lstick{data}&\qw&\qw&\multigate{1}{\mathcal S_Z}&\gate{\mathcal{R}}&\multigate{3}{\mathcal S_{Z|X}}&\gate{\mathcal R}&\multigate{5}{\mathcal S_X}&\gate{\mathcal R}&\qw\\
&&\gate{\mathcal{G}}&\multigate{1}{\mathcal{V}}&\ghost{\mathcal S_Z}&\control\cw\cwx[-1]\\
&&&\ghost{\mathcal{V}}&\control\cw\cwx\\
&&\gate{\mathcal{G}}&\multigate{1}{\mathcal{V}}&\qw&\qw&\ghost{\mathcal S_{Z|X}}&\control\cw\cwx[-3]\\
&&&\ghost{\mathcal{V}}&\cw&\cw&\control\cw\cwx[-1]\\
&&\gate{\mathcal{G}}&\multigate{1}{\mathcal{V}}&\qw&\qw&\qw&\qw&\ghost{\mathcal S_X}&\control\cw\cwx[-5]\\
&&&\ghost{\mathcal{V}}&\cw&\cw&\cw&\cw&\control\cw\cwx[-1]\\
}}}
\vspace*{13pt}
\fcaption{\label{fig:Steane_QEC}The error-correction network (1-EC) based on Steane's $[[7,1,3]]$ code, performing first {\tt Z}-error correction, and
then {\tt X}-error correction. Error correction consists of ancilla
preparation ($\mathcal{G}$), ancilla verification ($\mathcal{V}$),
syndrome extraction ($\mathcal{S}$), and a recovery operation
($\mathcal{R}$).  In many circumstances, the application of the
corrective gate $\mathcal{R}$ can be postponed and is carried forward classically
as a change of `Pauli frame' \cite{knill:nature}. }
\end{figure}

The local versions of ${\cal G}$ and ${\cal V}$ are shown in
Fig.~\ref{fig:latticeG}. Our ancilla verification scheme is less extensive
than the standard CSS method described, for example, in
Ref.~\cite{AGP:ft}. In the standard method, the ancilla is verified
using another block in the state $\ket{\bar{0}}$. For the
fault tolerance of our error-correction network, we only need to
measure the single stabilizer ${\tt IZIZIZI}$ on the ancilla block
$\ket{\bar{0}}$ \footnote{Thanks to Panos Aliferis for suggesting
this simplification.}. The ancilla needs to be verified since the
encoding uses two-qubit gates which can put two errors on the block.
If these errors are {\tt X} errors they can propagate to the data.
However, not every possible pair of {\tt X} errors will be generated
by the encoder circuit. In fact, we only need to check for errors on
the second, fourth, and sixth qubits of the ancilla block. Our
verification method prevents an ancilla from passing with the errors
${\tt X}_2 {\tt X}_7$, ${\tt X}_3 {\tt X}_6$, ${\tt X}_4 {\tt X}_5$,
each of which can be caused by a single faulty location in the
preparation circuit.  An ancilla with single {\tt X} and/or single
{\tt Z} errors can still pass verification.  To parallelize the
necessary measurement, verification is performed by preparing a
3-qubit cat state $(\ket{000}+\ket{111})/\sqrt{2}$ and applying
controlled-${\tt Z}$ ({\tt CZ}) gates between the three verification
qubits and the second, fourth, and sixth qubits of the ancilla
block. Each verification qubit is then measured in the {\tt
X} basis. If the parity of the measurement results is $1$, the
ancilla block fails verification. If an ancilla block passes
verification, it is used for syndrome extraction ($\mathcal{S}$),
shown in Figure \ref{fig:S}. During ancilla preparation the data
qubits are moved to home base in order to interact with the prepared
ancilla.


\setlength{\unitlength}{1cm}
\begin{figure}[h]
\centerline{
    \begin{picture}(13,9.5)
    \epsfig{file=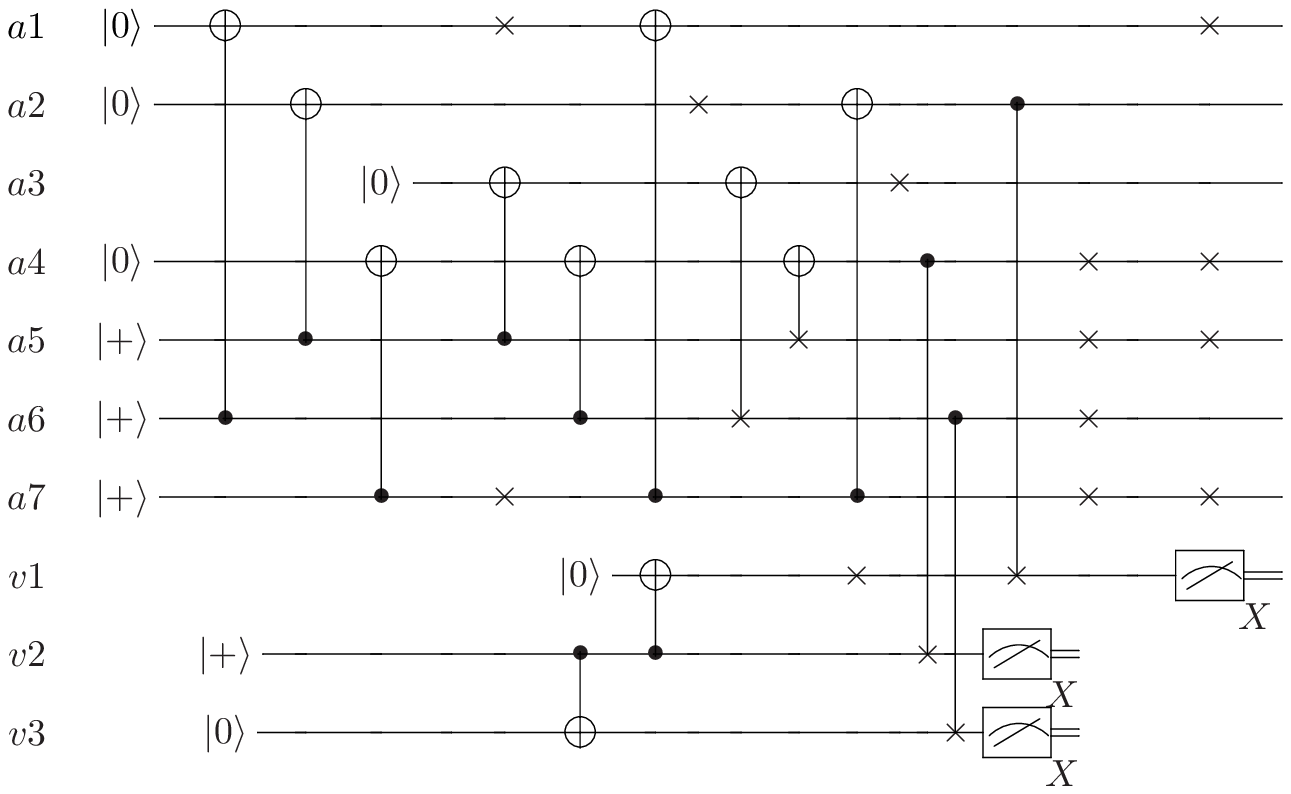,width=12.5cm}
    \put(-10.8,-0.1){\line(0,1){9}}
    \put(-8.2,-0.1){\line(0,1){9}}
    \put(-6.4,-0.1){\line(0,1){9}}
    \put(-4.4,-0.1){\line(0,1){9}}
    \put(-3.1,-0.1){\line(0,1){9}}
    \put(-1.3,-0.1){\line(0,1){9}}
    \end{picture}
}
\vspace*{13pt}
\fcaption{\label{fig:latticeG}The local preparation and verification networks
$\mathcal{G}$ and $\mathcal{V}$. The lower three qubits are the
verification qubits of network ${\cal V}$.  The seven timesteps in
which this circuit is executed are delineated by the long vertical
lines.  The third timestep is shown in Fig.~\protect\ref{fig:lattice}.  The single-qubit $\times$ symbol
indicates that in this timestep, the qubit has been swapped with a
dummy qubit. The \CNOT\ symbol with the control indicated with a
$\times$ means a \CNOT\ followed by \SWAP\ on the same qubits. The
two-qubit gate indicated by a solid circle on one end and a $\times$
on the other is a \CZ\ followed by {\tt SWAP}. It is important to
note that the qubit labels on the left correspond to the associated
wires throughout the circuit (and thus are never swapped) and thus
qubit a6, for example, interacts with qubit v3 in the 5th timestep. Thus the
fourth qubit travels the farthest, being swapped four spaces along
the square lattice during the operation of this circuit.}
\end{figure}

\begin{figure}[htb]
\centerline{ \mbox{
\subfigure[$\mathcal{S}_X$]{\scalebox{1.0}{\includegraphics{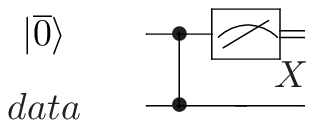}}}
\quad
\subfigure[$\mathcal{S}_Z$]{\scalebox{1.0}{\includegraphics{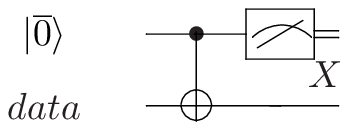}}}
} }
\vspace*{13pt}
\fcaption{\label{fig:S}The network $\mathcal{S}$ for syndrome extraction
during (a) {\tt X}-error correction and (b) {\tt
Z}-error correction. During syndrome extraction the data qubits sit
at home base in the cell. The two-qubit gate with solid
circles on both ends is the {\tt CZ} gate.}
\end{figure}


\subsection{Local Preparations (1-Preps)}\label{sec:local1prep}
\noindent The local 1-Prep for $\ket{\bar{0}}$ consists of at most
two attempts to create a verified ancilla as in the local network
in Fig.~\ref{fig:latticeG}.  We obtain the 1-Prep for
$\ket{\bar{+}}$ by applying transversal Hadamards on the passed
ancilla. For universality, we also need 1-Preps for the state
$\ket{\overline{a}_{\theta}}$, for $\theta=\pi/2,\pi/4$ (see
Fig.~\ref{fig:genST}). We can produce such states as follows (see
AGP~\cite{AGP:ft}: The unencoded state $\ket{a_{\pi/2}}$ is
correctly generated by measuring the operator ${\tt S} {\tt X}
{\tt S}^{\dagger}$ on the state $\ket{0}$, since ${\tt S}{\tt
X}{\tt S}^\dagger \ket{a_{\pi/2}} = \ket{a_{\pi/2}}$.  If we find
eigenvalue $+1$, we are done, otherwise we apply {\tt Z}.
Similarly, for $\ket{a_{\pi/4}}$, we measure ${\tt T} {\tt X} {\tt
T}^{\dagger}$ and correct with {\tt Z} if eigenvalue $-1$ is
found.  As we will see later, $\ket{\overline{a}_{\theta}}$
preparations are fundamentally different from Clifford-group
preparations, in that it is not necessary to make them fault
tolerant at every level of concatenation.

\subsection{Local Rectangles and Their Use of Space and Time}
\label{sec:spacetime}
\noindent

\begin{table}[htb]
\tcaption{\label{table:1ECloc}Types and number of locations
contained in a local 1-EC and two local 1-Gas. The first numbers
for the 1-EC are based on the assumption that the first and second
ancillas pass verification; the numbers in parentheses result from
the first or second ancilla failing verification. The last column
lists the number of locations in a horizontal \CNOT\ 1-Ga.}
\centerline{\footnotesize\smalllineskip
\begin{tabular}{|c|c|c|c|c|c|}
\hline
$\ell$ & Description & $\#$ in 1-EC & $\#$ in h-{\tt CNOT} 1-Ga \\
\hline
\hline 0 & horizontal \CNOT\ (h-{\tt CNOT}) & 2 (3) & 0 \\
\hline 1 & vertical \CNOT\ (v-{\tt CNOT}) & 30 (38) & 7 \\
\hline 2 & horizontal \CZ\ (h-{\tt CZ}) & 0 & 0  \\
\hline 3& vertical \CZ\ (v-{\tt CZ}) & 14 (14) & 0 \\
\hline 4& h-\CNOT\ followed by \SWAP\ & 0& 0 \\
\hline 5& v-\CNOT\ followed by \SWAP\  & 0& 0 \\
\hline 6& h-\CZ\ followed by \SWAP\ & 6 (9)& 0 \\
\hline 7& v-\CZ\ followed by \SWAP\  & 0& 0 \\
\hline 8& {\tt H}  & 0& 0 \\
\hline 9& horizontal \SWAP\ (h-{\tt SWAP})  & 22 (33)& 112  \\
\hline 10& vertical \SWAP\ (v-{\tt SWAP}) & 8 (12)& 14  \\
\hline 11& $\ket{+}$-preparation & 8 (12)& 0 \\
\hline 12& $\ket{0}$-preparation & 12 (18) &0  \\
\hline 13& {\tt X}-measurement & 34 (37)&0  \\
\hline 14& {\tt Z}-measurement & 0& 0 \\
\hline 15& wait during gate/prep & 18 (27) & 196 \\
\hline 16& wait during measurement & 0& 14  \\
\hline
\end{tabular}
}
\end{table}

In Table \ref{table:1ECloc}, we list the universal set of location types considered in our scheme.
Preparation of $\ket{+}$, measurement in the {\tt X} basis, the \CZ\
gate, and \CNOT\ and \CZ\ combined with \SWAP\ are considered single
locations which can be implemented in one fundamental timestep. We will assume that these composite gates have
error probabilities similar to the other elementary
gates. In practical implementations, some gates may have to be
constructed from multiple gates and may thus have higher
error rates. We assume that the time it takes to do a measurement is
the same as the single-qubit gate time and that classical
post-processing does not take any additional time.
Note in Table \ref{table:1ECloc} that we use two different types
of wait locations, one that happens in parallel with a gate and
one that happens in parallel with a measurement. The reason for
having separate locations is that the encoded version of the wait
location that occurs in parallel with a gate needs many additional wait locations in
order to be synchronized with the encoded version of a gate.

We include both horizontal and vertical two-qubit gate locations.
Two-qubit gates occurring between two logical qubits $x$ and $y$ on
a horizontal axis use the horizontal channel internal to lattice
cell $x$ or lattice cell $y$ as well as the horizontal channel
internal to the lattice cell belonging to the neighboring qubit
directly south of lattice cell $x$ or lattice cell $y$ (i.e.,
directly below it). We will actually use {\em either} the channel
south of lattice cell $x$ {\em or} the channel south of lattice cell
$y$, but not both. For example, in Fig.~\ref{fig:lattice}(a), data
qubits $d6$, $d5$, and $d3$ move into the channel internal to their
lattice cell, while qubits $d4$, $d2$, $d1$, and $d7$ move into the
neighboring channel directly south. To perform a two-qubit gate, the
neighboring horizontal qubits then move across the lattice cell by
swapping (for example, along row 2 and 6 in the figure) while the
qubits that moved into the channels remain stationary and wait.
Vertical two-qubit gates only use the vertical channel internal to
the lattice cells of the two qubits. \indent It follows that two
vertical two-qubit gates on two adjacent pairs of cells can be
executed in parallel since there is no common use of channels. For
two horizontal two-qubit gates on pairs of cells that are above each
other we need some scheduling since a common channel is used. This
scheduling is easy to achieve, since we can designate which of the
two qubits involved in the two-qubit gates will move into the
channel, and which will remain within the cell. The qubits remaining
within the cell can swap across the internal row freely, while the
qubits moving into the channel remain stationary and wait until the
data qubits moving along the interior of the lattice cell arrive.
For a more detailed explanation, a complete snapshot of the movement
just described is available at

\smallskip
http://www.research.ibm.com/quantuminfo/svore/movies.html .
\smallskip

In the local model, we must clearly define to what rectangle some
of the \SWAP\ operations and additional wait locations belong since
this can be ambiguous. During ancilla preparation in the 1-EC, the
data qubits move around in the cell and reside at home base once
syndrome extraction begins. This movement and possibly waiting of
data qubits is included in the definition of the previous 1-Ga,
since it can be viewed as movement back to home base after the
1-Ga was performed. Similarly, the 1-EC does not contain any \SWAP\
or wait locations on the {\em data} once the last syndrome
extraction (and possibly error correction) has been performed. Thus
the 1-Ga includes all data locations between the last time the
data was in home base for syndrome extraction and the next time the
data is in home base for syndrome extraction. The 1-EC includes all
locations on the data between the first and last syndrome extraction, including these syndrome
extractions (and possibly error correction) themselves.

With these definitions, we list the number of locations in a few essential routines in Table \ref{table:1ECloc}.
Note that the total number of {\tt SWAP} gates in the h-{\tt CNOT} 1-Ga is $112+14=126$ which is substantial.
As we will see in Section \ref{sec:results} the effect on the threshold of these additional gates is quite small.

In Table \ref{table:recsize}, we list the number of timesteps and locations
in a 1-Rec and 1-exRec for different locations $\ell$.
The local 1-EC takes 27 timesteps; this may be compared to the same
1-EC without {\tt SWAP}, i.e., the nonlocal 1-EC, which takes 21 timesteps.
Note that in Table \ref{table:recsize} the difference between the number of timesteps
in a 1-Rec and a 1-exRec is 26, which is one less than the number of timesteps in a 1-EC.
This is because there is an overlap of one timestep between the leading 1-EC and the 1-Rec.

Of all gates the horizontal \CNOT\ 1-Rec takes
the longest, a total of 35 timesteps (compared to 22 timesteps in
the similar nonlocal model). The single-qubit 1-Rec takes 28 timesteps
(compared to 22 timesteps in the nonlocal model). For
synchronization, we have added wait locations to the 1-Recs that take
fewer than 35 timesteps, so that the time for all `gate' 1-Recs is 35 timesteps.
Notice that the preparation 1-Recs take more time, but these 1-Recs can be started at the appropriate time
in advance so that the prepared states will be ready when necessary.

Since it is hard to represent what happens in three-dimensional
space-time on a two-dimensional sheet of paper, we only partially
represent in this paper the local fault-tolerant circuits that we
have developed. Supplementary material in the form of complete
sequences of time snapshots of the 2D layouts can be found at

\smallskip
http://www.research.ibm.com/quantuminfo/svore/movies.html.
\smallskip


\begin{table}[htb]
\tcaption{\label{table:recsize}The number of timesteps in the 1-Rec and the 1-exRec for different locations $\ell$.}
\centerline{\footnotesize\smalllineskip
\begin{tabular}{|c|c|c|c|}
\hline
$\ell$ & Time of 1-Rec & Time of 1-exRec & Total $\#$ loc. in 1-exRec \\
\hline\hline [0--7] & 35 & 61 & 1225 \\
\hline [8--10] & 35 &
61 & 616\\
\hline [11--12] & 41 & 41 & 469\\
\hline [13--14] & 1 & 27 & 196\\
\hline [15] & 35 & 61 & 616\\
\hline [16] & 28 & 54 & 378\\
\hline
\end{tabular}
}
\end{table}

\section{Methodology}
\label{sec:method}
\noindent
Our threshold analysis relies on counting the number of
malignant fault pairs in a 1-exRec; thus it is important to
understand this notion. In principle, we have to assume an
arbitrary input to the 1-exRec in order to analyze malignancy of faults. Given the fact that the [[7,1,3]] code
is a perfect code, incoming errors can be modeled by letting the input block have
at most 1 {\tt X} and 1 {\tt Z} error. Now one considers two Pauli
faults in the 1-exRec. If both faults occur in a single leading
1-EC, the output of that 1-EC can again be viewed as some codeword
plus at most one {\tt X} and {\tt Z} error. This implies that the
subsequent 1-Rec is correct, and thus the fault-pattern is benign.
If the 1-exRec is a \CNOT\ 1-exRec, both leading 1-ECs can have
faults and the transversal \CNOT\ gate could spread these so that
the 1-Rec may not be correct.

Let us consider how the possible input errors affect the output of a
1-EC with at most one fault. If the 1-EC has no fault, the incoming
errors will be corrected and the output is a codeword, just as if
there were no incoming errors. If the 1-EC has a single fault, it is
guaranteed that one of the incoming errors ({\tt X} or {\tt Z}) will
be corrected. As was argued in Ref.~\cite{AGP:ft}, the overall
effect of the other incoming error and the fault in the 1-EC is a
possible logical operation in the code space plus an error whose
position and character only depends on the fault in the 1-EC. In
other words, the errors at the end of the leading 1-EC with at most
one fault do not depend on the pattern of incoming errors. This
property still holds with our modified 1-EC where the 1-EC can
create one {\tt Z} and one {\tt X} error in the outgoing data block.
This implies that for determining malignancy, we only need to place
two faults inside the 1-exRec and not consider incoming errors.
Since we are using an adversarial noise model, we consider a pair of
locations malignant if it is malignant for some Pauli errors
occurring at those locations.

\subsection{Calculation of Failure Probabilities}
\label{sec:calcfail}
\noindent
We calculate the failure probability $\gamma_\ell^r = {\bf
F}_\ell({\bf \gamma}^{r-1})$ of a type-$\ell$ 1-exRec. Note that
${\bf F}$ is level-independent and we drop the superscript
$r$ in what follows. We can bound the failure probability
$\gamma_\ell$ as
\begin{eqnarray}
P(\mbox{failure of a 1-exRec}) & = & P(\mbox{1-Rec is incorrect}) \nonumber \\
&  \leq &   P(1^+ \mbox{ trailing 1-ECs do not occur})\nonumber\\
&  & +\, P(\mbox{1-Rec is incorrect {\tt \&} trailing 1-ECs occur}).
\label{eq:failure}
\end{eqnarray}
Remember that a 1-EC does not occur when less than two ancilla blocks
pass verification. Consider $P(\mbox{1-Rec is incorrect {\tt \&}
trailing 1-ECs occur})$. Because of the fault tolerance of the
quantum circuits, the 1-Rec is only incorrect when at least two
malignant faults occur in the 1-exRec. We can upper-bound this
probability as \bea P(\mbox{1-Rec is incorrect {\tt \&} trailing
1-ECs occur}) & \leq &
P(\mbox{1-Rec incor. due to mal. pair} \nonumber \\
& & \mbox{{\tt \&}\ trailing 1-ECs occur}) \nonumber \\
& & +\ P(3^+ \mbox{ faults {\tt \&} trailing 1-ECs occur})
\label{eq:2+fail} \eea
Consider the failure probability due to malignant pairs. In this
probability it is not specified whether the leading 1-ECs in the
1-exRec occurred or not. But the only way for a leading 1-EC to fail
is when there are two faults in it. As argued earlier in this section,
for the [[7,1,3]] code it follows that the 1-Rec is correct and
thus we can safely assume that the leading 1-ECs always occur. We
estimate this probability using a numerical malignancy count
\bea P(\mbox{1-Rec incor. due to mal. pair in 1-exRec of type
$\ell$}\nonumber \\
\mbox{{\tt \&} all 1-ECs occur in 1-exRec}) \leq \sum_{i \geq j}
\alpha_{[i],[j]}^\ell \gamma_i \gamma_j. \eea
Here $\alpha_{[i],[j]}^{\ell}$ is the number of malignant pairs of type $i$ and
$j$ for a 1-exRec of type $\ell$, where we count only those malignant pairs that do not lead to failure of a 1-EC.
The numerical malignant pair counts are obtained using the QASM
Toolsuite \cite{C:toolsuite}. With this software, one simulates
the propagation of Pauli errors in a quantum circuit and thus
determines whether a pair of faults is malignant. A selection of these
malignancy matrices $\alpha_{[i],[j]}^{\ell}$ are reproduced in
Appendix \ref{app:alphas}.

We upper-bound the triple-fault term as $P(3^+ \mbox{ faults {\tt \&}
trailing 1-ECs occur}) \leq {N \choose 3} \gamma_{\rm{max}}^3,$
where $N$ is the total number of locations in the full 1-exRec
(including all 1-ECs) and $\gamma_{\rm{max}} = \mbox{max}_{\ell}({\bf
\gamma}_{\ell})$ is the maximum failure probability. This
expression is probably a large over-estimate of the higher-order
terms. For our largest local 1-exRec, the \CNOT\ 1-exRec, $N=1225$.
This implies that, including {\em only} these triple malignancies,
the threshold cannot be above $p=5.7 \times 10^{-5}$ (which is the
solution of the equation $p=p^3 {1225 \choose 3}$). Since the third
term is a very steep function of $p$, the effect of the term around
the thresholds that we find below is not so large. In the results in
Section \ref{sec:results}, we see that the effect of omitting
this term altogether gives similar thresholds in the $O(10^{-5})$
range. Alternative methods of bounding this higher-order term exist,
but they either involve triple malignancy counting (which is very
time consuming) or heuristic strategies such as malignancy sampling.

Let us return to the first term in Eq (\ref{eq:failure}). We
calculate the probability that one or more of the 1-ECs fail as
\begin{eqnarray}
P(1^+\mbox{trailing 1-ECs fail}) = 1 - P(\mbox{1-EC occurs})^{q},
\end{eqnarray}
where $q=0$ for a measurement 1-exRec, $q=1$ for single-qubit 1-exRecs (including \SWAP\ and 1-Prep), and $q=2$
for the two-qubit 1-exRecs. Furthermore,
\begin{eqnarray}
P(\mbox{1-EC occurs}) = P(\mbox{anc. pass})^2 + 2P(\mbox{anc.
pass})^2P(\mbox{anc. fail}). \label{eq:1-ECoccur}
\end{eqnarray}

How do we estimate the probability for an ancilla to pass?
We have $P(\mbox{anc. pass})  = 1 - P(\mbox{anc. does not pass})$,
where we bound \beq P(\mbox{anc. does not pass}) \leq P(\mbox{no pass due to 1 fault in }
\mathcal{G},\mathcal{V}) + P(2^+ \mbox{ faulty loc. in }
\mathcal{G},\mathcal{V}),
\eeq
where $P(2^+ \mbox{ faulty loc. in } \mathcal{G},\mathcal{V}) \leq {49 \choose 2} \gamma_{\rm{max}}^2$.
Thus we assume any two faulty locations in preparation and verification cause
the ancilla to fail verification. This is a slight overcount, but the majority of single
faulty locations in fact cause the ancilla to fail verification. We have
\begin{eqnarray}
P(\mbox{no pass due to 1 fault in } \mathcal{G},\mathcal{V}) =
\sum_{\ell \in \mathcal{G},\mathcal{V}} \alpha_\ell \gamma_\ell,
\end{eqnarray}
where $\alpha_\ell$ is the number of bad locations of type $\ell$
occurring in the $\mathcal{G},\mathcal{V}$ network as determined
by the QASM toolsuite, listed in Table \ref{table:malloc}.

\begin{table}
\tcaption{\label{table:malloc}Number of bad locations of type $\ell$ in the
$\mathcal{G}$ and $\mathcal{V}$ networks that make the ancilla not
pass verification. Location types not listed are all good.}
\centerline{\footnotesize\smalllineskip
\begin{tabular}{|c|c|}
\hline
Description of type $\ell$ & Number\\
\hline \hline h-\CNOT\ & 1 \\
\hline v-\CNOT\  & 7 \\
\hline h-\CNOT\ followed by \SWAP\ & 2 \\
\hline h-\CZ\ followed by \SWAP\ & 2 \\
\hline h-\SWAP\  & 3 \\
\hline $\ket{0}$-preparation & 2 \\
\hline $\ket{+}$-preparation & 1 \\
\hline {\tt X}-measurement  & 3 \\
\hline wait during gate/prep & 2 \\
\hline
\end{tabular}
}
\end{table}

\section{Threshold Studies}\label{sec:results}
\noindent We determine a lower bound on the noise threshold for
each location type using the formulas for failure probabilities
given in the previous sections. In order to estimate thresholds,
we repeatedly apply the 17-dimensional flow map ${\bf F}_\ell({\bf
\gamma})$ to the initial vector of failure probabilities. The only
flow that depends on the $a_{\pi/2}$- and $a_{\pi/4}$-preparations
is the flow map for these preparations themselves. This implies
that the threshold for all the other (Clifford) gates is {\em
independent} of the initial failure probabilities for these two
preparations. Thus we can estimate the threshold for these
preparations separately and see whether or not they determine the
overall threshold.

\subsection{Clifford-group Location Thresholds}
\noindent
Let us first consider a much studied case in the literature in
which all initial probabilities are the same, i.e., set to some value
$\gamma$, except the failure probability for a memory location
which is set to $\gamma/10$. The threshold
of all these gates (except the memory location) will be denoted as
$\gamma_c$. For our 2D lattice architecture, we find a lower bound of
$\gamma_{c} \geq 1.85 \times 10^{-5}$.  We compare our local 2D
architecture with a nonlocal architecture with exactly the same
circuits (except for the {\tt SWAP}s).  For the nonlocal setting,
we find $\gamma_c \geq 3.61 \times 10^{-5}$.

In Section \ref{sec:calcfail} we argued that we considerably overestimate the
three-or-more-fault terms in our failure probability calculation. To see how this affects
the threshold, we perform our analysis omitting this triple-fault term entirely in Eq (\ref{eq:2+fail}).
The local threshold then becomes $2.15 \times 10^{-5}$ and the nonlocal threshold becomes $4.32 \times 10^{-5}$, which
are small corrections on the numbers given above.

To test the correctness of our local analysis, we also analyze our
local 2D architecture in a variant where at all levels of
concatenation the error rates of \SWAP\ gates are set to zero by
hand. Then the only difference between the local and nonlocal case
is the additional wait locations in the local architecture. In this
scenario, the threshold is $\gamma_c \geq 3.0 \times 10^{-5}$, in
which we see the effect of additional waiting in the local model.
To eliminate the possibility that the threshold for the nonlocal
architecture is suboptimal due to the choice of deterministic
error correction, we also analyze the nonlocal circuits in
\cite{AGP:ft}; this gives a lower bound of $\gamma_c \geq 4.3 \times
10^{-5}$. This shows that our choice of circuits is close to
optimal.

One may think that the closeness of the local and nonlocal
thresholds could be due to the fact that we let the memory
error rates be $1/10$ of the gate error rates and thus eliminate the
effect of additional waiting on the local threshold. Thus we also
consider the local and nonlocal architecture where all error
probabilities are identical. For the local architecture we obtain a
lower threshold, $\gamma_c \geq 1.1 \times 10^{-5}$; for the
nonlocal architecture the threshold becomes $1.9 \times 10^{-5}$.

\begin{figure}[htb]
\centerline{
\includegraphics[totalheight=0.25\textheight]{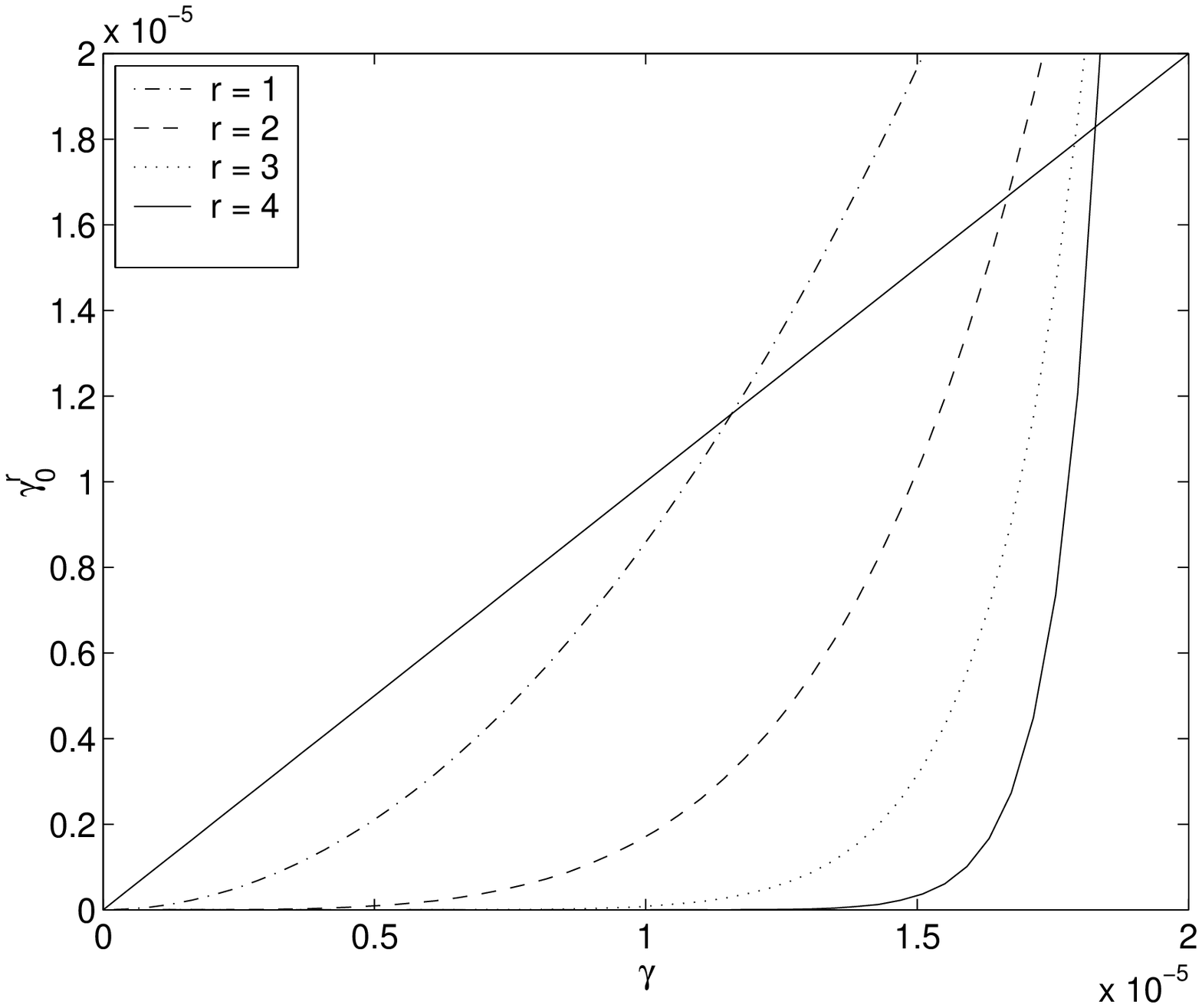}}
\vspace*{13pt}
\fcaption{\label{fig:tenth}Initial failure probability $\gamma$ versus the failure probability $\gamma_0^r$ at level $r$ of a horizontal \CNOT\ location,
shown for levels $r=1,\ldots,4$.}
\end{figure}

\begin{figure}[htb]
\centerline{
\includegraphics[totalheight=0.25\textheight]{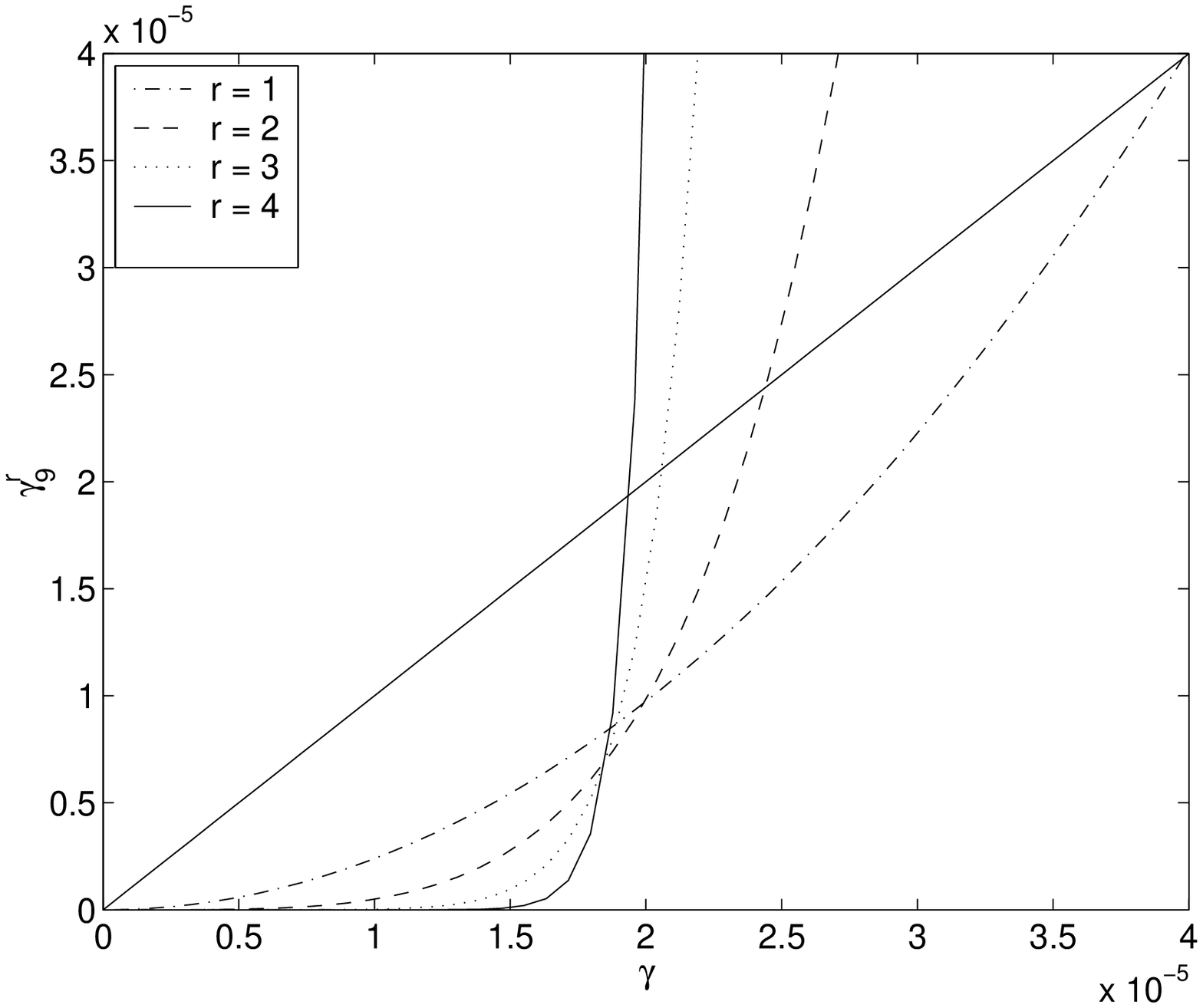}}
\vspace*{13pt}
\fcaption{\label{fig:tenthswap}Initial failure probability $\gamma$ versus the failure probability at level $r$ of a horizontal \SWAP\ location type
$\gamma^r_9$, shown for levels $r=1,\ldots,4$.}
\end{figure}

To understand the behavior of the failure probabilities at different levels
of encoding we use a {\it threshold reliability information plot}
(TRIP) \cite{SCCA:flowmap}. The behavior of the failure probability
for the horizontal \CNOT\ in the local model is given in
Fig.~\ref{fig:tenth}. The value of $\gamma$ at the crossing point
with the 45-degree line is the threshold at level $r$, also called the level $r$ {\em pseudothreshold},
for the given location type. In Figure \ref{fig:tenthswap}, we have plotted the
behavior of the \SWAP\ location. From these figures it is clear that
the horizontal \CNOT\ dominates the threshold. The pseudothresholds
of the \SWAP\ location get worse as a function of $r$ due to the noisiness of the {\tt CNOT}, whereas the
pseudothresholds of the \CNOT\ get better as a function of $r$. Another way of seeing this is that
after one level of encoding there is a region in which one is below the pseudothreshold of the
\SWAP\ gate, but not below the pseudothreshold of the \CNOT\ gate. In this region, the \CNOT\ gets
worse, which feeds into a worse pseudothreshold of the \SWAP\ at the second level of encoding.

\subsection{Approximate Closed Form for Thresholds}\label{sec:closedform}
\noindent Even though our map is high-dimensional, we can capture
the essential features by a two-dimensional map.  We do this by
making some drastic approximations: the 1-EC failure probabilities
of Eq. (\ref{eq:failure}), and the 3+ fault probabilities of Eq.
(\ref{eq:2+fail}), are ignored.  Many distinct rectangles are
treated as having the same error probabilities.  Despite these
approximations, the thresholds can be estimated very simply to
about a factor of two.

We assume that all error levels of gates, preparations, and
measurements are comparable (but not necessarily identical) at the
base level. If some of the error rates are much larger than others
(say, by a factor of 100), it may take more levels of
concatenation for the two-dimensional representation to become a
good approximation, but these initial conditions will be
eventually marginalized.

We only consider the Clifford gates. After a single iteration, information
about the locations in each 1-exRec is absorbed into the new error
probabilities of simulated locations. Then we find if we start with comparable error rates for all gates,
 all subsequent iterations have the property that the failure
probabilities fall into three distinct groups: (1) those of
two-qubit gates (locations [0--7]) (2) those of single-qubit gates
(locations [8--10], [15], [16]), (3) those of the preparations (locations
[11], [12]) and those of measurement (locations [13], [14]). In other words,
the failure probabilities of members of each group are very close in
value.  We see that this grouping basically corresponds to the number of 1-ECs in the
1-exRec.

We also observe that the measurement failure probabilities are very
small compared to the other failure probabilities and that the behavior of the map
changes very little if the measurement failure probability is always
set to zero. After the second iteration, the failure probability of
the preparation rectangles gets close to those of the single-qubit
gates, so we group them together. Thus, we end up with a
two-dimensional map described by the equations:
\begin{equation}
x^{i+1}=\left(x^{i}\,\,y^{i}\right)\cdot\left(\begin{array}{cc}a&b\\b&c\end{array}\right)
\cdot\left(\begin{array}{c}x^{i}\\y^{i}\end{array}\right),
\end{equation}
\begin{equation}
y^{i+1}=\left(x^{i}\,\,y^{i}\right)\cdot\left(\begin{array}{cc}d&e\\e&f\end{array}\right)
\cdot\left(\begin{array}{c}x^{i}\\y^{i}\end{array}\right),
\end{equation}
where $i$ is the iteration level. Here $x^{i+1}$ is the failure
probability of the two-qubit gate group at level $i+1$ and $y^{i+1}$
is the failure probability of the single-qubit gate group. The numbers
$a$, $b$, etc. can be extracted by summing entries in the malignancy
matrices $\alpha_{[i],[j]}^{\ell}$ (see Appendix \ref{app:alphas} for two examples of single-qubit and
two-qubit gates). For the local model, the numbers for $\ell=0$ (horizontal {\tt CNOT}) and
$\ell=9$ (horizontal {\tt SWAP}) are $a=7907$, $b=\frac{55997}{2}$ and $c=93488$
and $d=1956$, $e=\frac{18424}{2}$ and $f=35886$.

The equations for the fixed point of this map are two coupled
second-order equations for $x$ and $y$.  By elementary elimination
methods, one can derive from these a fourth-order polynomial in $x$
whose roots give the fixed point.  One root is at $x=0$; factoring
this out leaves a cubic polynomial.  We find that, for the range of
parameters we have, the next root, giving the threshold for $x$, is
accurately given by truncating this cubic polynomial to linear
order.  The resulting expression for the threshold value of $x$ is
\begin{equation}
x={c\over ac+4ce-2bf-f^2}. \label{quad1}\end{equation} A parallel
analysis for $y$ gives an expression for its fixed point:
\begin{equation}
y={d\over a^2+4bd-2ae+df}. \label{quad2}\end{equation}

Now we analyze the first iteration of the full map. For simplicity, we try to follow the
analysis in the previous section where all error rates are the same, except for the memory error rates.
At the physical level, we start with a gate failure rate $\gamma_g$ and a
wait failure rate $\gamma_w$. The $x^{1}$ and $y^{1}$ are given by
quadratic expressions
\begin{equation}
x^{1}=\left(\gamma_g\,\,\gamma_w\right)\cdot\left(\begin{array}{cc}F&G\\G&H\end{array}\right)
\cdot\left(\begin{array}{c}\gamma_g\\
\gamma_w\end{array}\right),
\end{equation}
\begin{equation}
y^{1}=\left(\gamma_g\,\,\gamma_w\right)\cdot\left(\begin{array}{cc}J&K\\K&L\end{array}\right)
\cdot\left(\begin{array}{c}\gamma_g\\
\gamma_w\end{array}\right).
\end{equation}
For the local model, the parameters are $F=71779$,
$G={77899\over 2}$, $H=26842$ (taken from the horizontal
{\tt CNOT} malignancy matrix) and $J=8318$, $K={32843\over 2}$ and
$L=21632$ (extracted from the single-qubit wait malignancy
matrix).

We have been interested in the case $\gamma_w={1\over
10}\gamma$.  We find this case is numerically similar to setting
$\gamma_w=0$, causing these equations to become very simple:
$x^{1}=F \gamma^2$, and $y^{1}=J\gamma^2$.  Setting
$x^{1}=x$, the threshold expression, and $y^{1}=y$ (the threshold
expressions Eqs (\ref{quad1},\ref{quad2}) and solving for
$\gamma$, we get two different expressions for $\gamma$:
\begin{equation}
\gamma=\sqrt{c\over F(ac+4ce-2bf-f^2)},
\end{equation}
\begin{equation}
\gamma=\sqrt{d\over J(a^2+4bd-2ae+df)}.
\end{equation} We will be below threshold if $\gamma$ is lower
than either of these two expressions, since then both $x^{1}$ and
$y^{1}$ will be below their threshold values.  Therefore,
\begin{equation}
\gamma_{c}=\min\left(\sqrt{c\over F(ac+4ce-2bf-f^2)}\,\,
,\,\, \sqrt{d\over J(a^2+4bd-2ae+df)}\,\,\right). \end{equation}

Note that these expressions satisfy the required homogeneity
property, namely that if all malignant pair counts are scaled by the same
constant $\kappa$, the threshold changes by the factor $1/\kappa$.
Putting in values for these malignant pair counts, we find that
the estimated threshold values are $\gamma_c \geq 1.94 \times
10^{-5}$ for the local model and $\gamma_c \geq  4.67 \times 10^{-5}$ for the
nonlocal model.  This agrees very well with the threshold values
of the actual approximate map, which are $1.94 \times 10^{-5}$ and
$4.43 \times 10^{-5}$; these are in turn in reasonable agreement
with the threshold values for the exact map, which were found
to be $1.85 \times 10^{-5}$ and $3.61 \times 10^{-5}$.

Our formulas explain the factor-of-two difference between the
local and nonlocal cases in the following way. The malignant pair
counts differ between the two models, in different cases by a
factor of 5 at most, and a factor of 1 (unchanged, that is) at the
least.  These formulas represent a kind of weighted average among
these different counts, so the actual difference is somewhere
between a factor of 5 and a factor of 1.

\subsection{$a_{\theta}$-preparation Threshold and Injection by Teleportation}
\label{sec:inj}
 \noindent It would be possible to approach
$a_{\theta}$-preparation in exactly the same way as all other
locations in the quantum circuit.  A deterministic, fault-tolerant
extended rectangle for this preparation is known, which is based
on the fact that the state $\ket{\bar{a}_{\theta}}$ is an
eigenstate of the encoded operator $\bar{R}_{\tt
Z}(\theta)\bar{\tt X}\bar{R}_{\tt Z}(-\theta)=\bar{R}_{\tt
Z}(2\theta)\bar{\tt X}$.  Fault tolerance is achieved by repeated
measurement of this operator followed by error correction, exactly
as in Fig. 13 of AGP~\cite{AGP:ft}. Since we require this
rectangle to be deterministic, we must allow for three
repetitions, rather than the two repetitions used in AGP.

Unfortunately, this triple-repetition exRec, with all the moves
and waits required by locality, has an extremely large number of
locations, and the threshold, if we determine it using the
formalism discussed for the Clifford gates, is very poor.  But we
can take advantage of the fact that the $\ket{{a}_{\theta}}$
preparations are not mandatory at the lowest level (because these
preparations are not needed in 1-ECs) to obtain fault tolerance by
a different strategy.

\begin{figure}[h]
\centerline{
\includegraphics{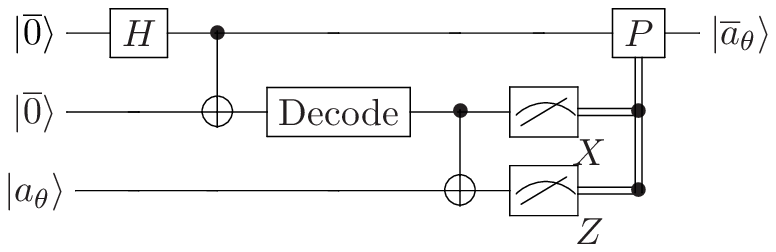}}
\vspace*{13pt} \fcaption{ \label{fig:tele}Injection by
teleportation. An encoded EPR state is created and half of the
state is decoded. A Bell measurement is carried out on the state
$\ket{a_{\theta}}$ and the decoded half followed by encoded
corrective Paulis on the encoded half of the EPR pair.  There is
ample room in the two-dimensional cell to lay out this circuit,
whatever the level of concatenation desired for the final state.}
\end{figure}

This strategy is {\em injection by teleportation}, an idea due to
Knill \cite{knill:nature}.  The key circuit is shown in
Fig.~\ref{fig:tele}; here, an encoded EPR pair is created and one
half of the pair is decoded.  This decoded half is used to
teleport an unencoded $\ket{a_{\theta}}$. After teleportation, the
resulting state is the encoded $\ket{\overline{a}_{\theta}}$. The
encoded state produced by this circuit may be at any level of
concatenation. This circuit is not fault-tolerant, so
$\ket{\overline{a}_{\theta}}$ may be noisier than
$\ket{a_{\theta}}$.  But this circuit will only suffer from single
faults in gates in the latter part of the decoding circuit and in
the measurements. The total number of such gates is on the order
of ten, no matter what the level of concatenation of the final,
coded state.  Thus, if we inject to level $k$ of teleportation,
the encoded error rate $\gamma_\theta^k$ is given approximately by
$\gamma_\theta^k=\gamma_\theta^0+10\gamma_{\mbox{\tiny CNOT}}^0$.

For purposes of analysis, we imagine that the injection level $k$
is chosen so that the Clifford-group gates are essentially
noiseless.  At this level, we may use a fault tolerant circuit
related to Fig. 13 of AGP~\cite{AGP:ft}, as described above.  Upon
further concatenation, the probability flows will be determined
just by the preparation counts in this 1-exRec.  There would be 84
$a_{\theta}$-preparations in the triple-repetition rectangle; the
$a_{\theta}$-threshold, then, is estimated by the equation
$\gamma_\theta={84\choose 2}\gamma_\theta^2$.  Assuming
$\gamma_{\mbox{\tiny CNOT}}=10^{-5}$, these equations give an
estimate for the $\gamma_\theta$ threshold of $1.0\times 10^{-4}$.

Thus, the $a_{\theta}$-threshold value is not unduly low; we
expect, in fact, that this number can be improved with
modifications of the AGP circuit.   We note that this threshold is
of course far from the theoretical upper maximum of about $17\%$
established by the analysis of the magic-state distillation
technique \cite{BK:magicdistill}.   We also note that, with the
scheme just outlined, there is a serious efficiency cost, in that
more levels of concatenation are typically needed to achieve
effectively noiseless operation.  We expect that this, too, may be
improved upon.

\section{Concluding Remarks on the Noise Threshold of the [[7,1,3]] Code}\label{sec:why10-5}
\noindent
In the fault-tolerance literature, various estimates of the nonlocal
threshold for the [[7,1,3]] code have been given. These numbers vary between
an optimistic value of $O(10^{-3})$ \cite{steane:overhead}, an estimate of $O(10^{-4})$
\cite{STD:localft} and the most recent rigorous {\em lower bound} of
$2.73 \times 10^{-5}$ established in Ref.~\cite{AGP:ft}. In this section,
we argue why, for stochastic independent noise models, the noise
threshold of the [[7,1,3]] code {\em is} realistically $O(10^{-5})$.

The main reasons that the earlier estimates were more optimistic
lies in the method of analysis. In Ref.~\cite{AGP:ft} one
estimates the failure probability of an {\em extended} rectangle,
which includes error correction before and after the encoded gate.
The previous work analyzed the failure probability of a rectangle
and tried to take into account incoming errors into the rectangle
in a heuristic fashion. The reason that the analysis in
Ref.~\cite{AGP:ft} is superior to the other analyses is that the
definition of failure introduced in Ref.~\cite{AGP:ft} has a
direct interpretation in terms of the failures in the unencoded
quantum circuit that the noisy quantum circuit simulates. Namely,
by pushing perfect decoders from the end of the computation
through to the front, one generates the unencoded circuit in which
failed 1-exRec corresponds to failed elementary gates.

In Ref.~\cite{AGP:ft}, the malignancy method was used to establish a
{\em lower bound} of $2.73 \times 10^{-5}$ for the $[[7,1,3]]$ code
in the presence of independent adversarial stochastic noise. This
derivation of the lower bound assumed, for simplicity, that all
gates in the circuit are (worst-case) \CNOT\ gates. One may thus
think that such an analysis would be far too pessimistic. This
conclusion is falsified by two observations. The first observation is that there are
already 13245 malignant \CNOT\ pairs {\em alone} in the \CNOT\
1-exRec (see the malignancy matrix in Ref.~\cite{AGP:ft}). Thus just assuming that all gates are noiseless except the
\CNOT\ would set a threshold of $\frac{1}{13245}
\approx 7.55 \times 10^{-5}$. The second observation is that the full nonlocal
analysis of the present paper that takes into account all types of
locations similarly produces a $O(10^{-5})$ threshold. The
threshold also does not change very much if we change adversarial
noise to depolarizing noise as shown in Ref.~\cite{AGP:ft}; their lower
bound for depolarizing noise is again $O(10^{-5})$.

Another cause for potential looseness in threshold estimates using
the malignancy method is how one deals with incoming errors in the
1-exRec. For general codes, the patterns of incoming errors into the
1-exRec may determine which pattern of faults is malignant inside
the 1-exRec. If we assume a worst-case pattern of incoming errors
(instead of some steady-state pattern), our threshold estimate could
be lower than necessary. But for the perfect $[[7,1,3]]$ code, one can
argue (as was done in Ref.~\cite{AGP:ft} and repeated here) that the
malignancy of faults inside the 1-exRec is independent of the
pattern of incoming errors.

These observations lead to two conclusions that are of interest in
further studies. Firstly, it is highly desirable to look at
different code architectures and see whether one can obtain a
threshold in the range $O(10^{-3})-O(10^{-4})$.  One example of such a
promising architecture is the $C_4/C_6$ scheme by Knill
\cite{knill:nature}. Secondly, as witnessed by the analysis of the
[[7,1,3]] code, thresholds have certain crucial dependencies but are
otherwise remarkably robust against small variations (in the encoding
circuitry, in the noise rates of some gates, and in adding {\tt SWAP}
gates for locality, for example). To identify these crucial
dependencies, by studying a whole range of codes and understanding in
what part of `code-space' the most promising codes lie, will
be an important task ahead.

\nonumsection{Acknowledgements}
\noindent
We would like to thank Panos Aliferis for many interesting
discussions and insightful comments. KMS would like to thank John
Preskill for the invitation to visit the Institute for Quantum
Information at Caltech. KMS also thanks Andrew Cross for his work on
the QASM toolsuite and also many engaging discussions. DPDV and BMT
acknowledge support by the NSA and the ARDA through ARO contract
number W911NF-04-C-0098. Some of our quantum circuit diagrams were
made using the Q-circuit \LaTeX\ macro package by Steve Flammia and
Bryan Eastin.


\nonumsection{References}

\appendix{\ Fault-tolerance Requirements}\label{app:ftprop}
\noindent
The $[[7,1,3]]$ code is a perfect code which means that all states in
the $2^7$ dimensional space can be represented as some codeword
with at most 1 {\tt X} error and at most 1 {\tt Z} error.
\begin{table}[htb]
\tcaption{\label{table:stab713}The six stabilizers and logical operations of the [[7,1,3]] code.}
\centerline{\footnotesize\smalllineskip
\begin{tabular}{|c|c|}
\hline
Name & Operator \\
\hline\hline $s_1$ & {\tt IIIXXXX} \\
\hline $s_2$ & {\tt IXXIIXX} \\
\hline $s_3$ & {\tt XIXIXIX} \\
\hline $s_4$ & {\tt IIIZZZZ}  \\
\hline $s_5$ & {\tt IZZIIZZ}  \\
\hline $s_6$ & {\tt ZIZIZIZ} \\
\hline & \\
$\bar{X}$ & {\tt XXXXXXX} \\
$\bar{Z}$ & {\tt ZZZZZZZ} \\
\hline
\end{tabular}}
\end{table}

The fault-tolerance properties that the constructions of 1-Gas,
1-Meas's, 1-Preps, and 1-ECs must satisfy for a distance-3 code, as
stated in Ref.~\cite{AGP:ft}, are:\\
1. If a 1-EC contains no fault, it takes any input to an output in the code space. \\
2. If a 1-EC contains one fault, it takes any input to a valid output. (The output of a level-1 block is `valid' if it deviates from the code space by the action of a weight-1 operator).\\
3. If a 1-EC contains no fault, it takes an input with at most one error to an output with no errors.\\
4. If a 1-EC contains at most one fault, it takes an input with no errors to an output with at most one error.\\
5. If a 1-Ga contains no fault, it takes an input with at most one error to an output with at most one error in each output block.\\
6. If a 1-Ga contains at most one fault, it takes an input with no errors to an output with at most one error in each output block.\\
7. A 1-Meas with no faults applied to an input with one error agrees with an ideal measurement. \\
8. A 1-Meas with at most one fault applied to an input with no errors agrees with an ideal measurement. \\
9. A 1-Prep with at most one fault produces an output with at most one error.

\appendix{\ Selected Malignant Pair Matrices}\label{app:alphas}
\noindent
In these matrices the columns and rows corresponding to location i are
indicated by [i]. Columns (and rows) with only zero elements are
omitted.

For the horizontal {\tt SWAP}, location 9, we have

\beq
\alpha_{[i],[j]}^9 = \ba{c} \, \\ \mbox{[0]} \\ \mbox{[1]} \\ \mbox{[3]} \\ \mbox{[4]}  \\ \mbox{[6]} \\ \mbox{[9]} \\
\mbox{[10]} \\ \mbox{[11]} \\ \mbox{[12]} \\ \mbox{[13]} \\ \mbox{[15]} \\
\mbox{[16]} \ea \left( \ba{ccccccccccccc}
[0] & [1] & [3] &[4]  & [6] & [9] & [10] & [11] & [12] & [13] & [15] & [16] \\
16&  186&    68&     40&         56&           592&    72&     42&     52&     106&       1043&   68 \\
  &   467&       312&    204&       276&          2856&   347&    200&     244&    480&        4931&   312 \\
  &    &       63&     70&         90&            942&    123&    90&      96&     210&        1782&   126 \\
  &    &        &   18&        60&            618&    78&     48&     60&      116&       1080&    70 \\
  &    &        &    &       30&            812&    106&    64&     76&     144&       1404&   90 \\
  &    &        &    &        &           3345&   1088&   716&    780&     1704&      10572&  606 \\
  &    &        &    &        &            &   60&      84&     100&     192&       1890&    123 \\
  &    &        &    &        &            &    &   12&     44&     72&        1338&   90 \\
  &    &        &    &        &            &    &    &   24&     96&        1398&   96 \\
  &    &        &    &        &            &    &    &    &   84&        3114&   210 \\
  &    &        &    &        &            &    &    &    &       &   11612&  1824 \\
  &    &        &    &        &            &    &    &    &       &    &   84 \\
\ea \right)\nonumber \eeq

For the horizontal {\tt CNOT}, location 0, we have \beq
\alpha_{[i],[j]}^0= \ba{c} \, \\ \mbox{[0]} \\ \mbox{[1]} \\ \mbox{[3]} \\ \mbox{[4]}  \\ \mbox{[6]} \\ \mbox{[9]} \\
\mbox{[10]} \\ \mbox{[11]} \\ \mbox{[12]} \\ \mbox{[13]} \\ \mbox{[15]} \\
\mbox{[16]} \ea \left( \ba{ccccccccccccc}
[0] & [1] & [3] &[4]  & [6] & [9] & [10] & [11] & [12] & [13] & [15] & [16] \\
60&  761&      241&    144&      198&        2018& 421&144 & 166&
364&      2567&   54 \\
  &  2106&     1295&   810&      1116&       10561&  2309&   806&    918&    1938&     13487&
 266 \\
  &   &    210&    250&      324&        3264&   692&    306&    312&    714&      4572&
 84 \\
  &   &     &  68&       216&        2103&   448&    164&    190&    392&      2668&
 56 \\
  &   &     &   &    108&        2799&   594&    224&    236&    504&      3496&
 72 \\
  &   &     &   &     &      11748&  5460&   2237&   2469&   5328&     29008&
 216 \\
  &   &     &   &     &       &  593&    508&    532&    1176&     6831&
 98 \\
  &   &     &   &     &       &   &  48&     152&    288&      3260&
 72 \\
  &   &     &   &     &       &   &   &  74&     336&      3268&
 72 \\
  &   &     &   &     &       &   &   &   &  336&      7584&
 168 \\
  &   &     &   &     &       &   &   &   &   &    25240&
 1560 \\
  &   &     &   &     &         &   &   &   &   &   &  42
 \\
\ea \right)\nonumber \eeq

\end{document}